# ISLAND: In-Silico Prediction of Proteins Binding Affinity Using Sequence Descriptors

Wajid Arshad Abbasi[a], Fahad Ul Hassan[b], Adiba Yaseen[a], and Fayyaz Ul Amir Afsar Minhas[a, †]

[a] *Biomedical Informatics Research Laboratory, Department of Computer and Information Sciences, Pakistan Institute of Engineering and Applied Sciences (PIEAS), Nilore, Islamabad, Pakistan*

[b] *Department of Electrical Engineering, Pakistan Institute of Engineer-ing and Applied Sciences (PIEAS), Nilore, Islamabad, 44000, Pakistan.*

[†]*Corresponding author's email*: afsar@pieas.edu.pk

**ABSTRACT**

Determination of binding affinity of proteins in the formation of protein complexes requires sophisticated, expensive and time-consuming experimentation which can be replaced with computational methods. Most computational prediction techniques require protein structures which limit their applicability to protein complexes with known structures. In this work, we explore sequence based protein binding affinity prediction using machine learning. Our paper highlights the fact that the generalization performance of even the state of the art sequence-only predictor of binding affinity is far from satisfactory and that the development of effective and practical methods in this domain is still an open problem. We also propose a novel sequence-only predictor of binding affinity called ISLAND which gives better accuracy than existing methods over the same validation set as well as on external independent test dataset. A cloud-based webserver implementation of ISLAND and its Python code are available at the URL: http://faculty.pieas.edu.pk/fayyaz/software.html#island.

## 1. Introduction

Protein binding affinity is a measure of the strength of the interaction between two binding proteins in a complex(Kessel and Ben-Tal, 2010). It is a key factor in enabling protein interactions and defining structure-function relationships that drive biological processes(Alberts et al., 2002). Accurate measurement of binding affinity is crucial in understanding complex biochemical pathways and to uncover protein interaction networks. It is also measured as part of drug discovery and design to improve drug specificity(Tomlinson, 2004). It can be measured in terms of dissociation constant ($K_d$) through different experimental methods such as Nuclear magnetic resonance spectroscopy, gel-shift and pull-down assays, analytical ultracentrifugation, Surface Plasmon Resonance (SPR), spectroscopic assays, etc (Kastritis and Bonvin, 2013; Wilkinson, 2004). However, the accuracy of these methods depends on dissociation rates and these methods cannot be applied at a large scale due to cost and time constraints (Vangone and Bonvin, 2015; Wilkinson, 2004). Therefore, accurate computational techniques can play an important role in affinity determination of protein complexes.

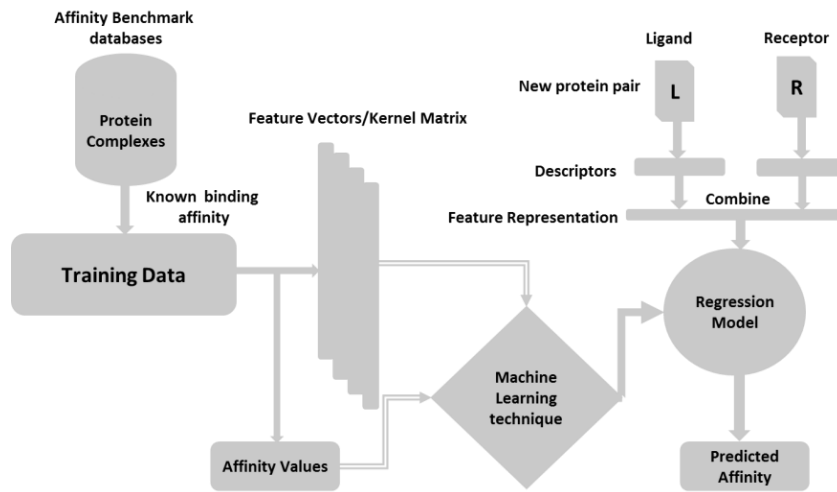

**Fig. 1.** A general framework of machine learning techniques for protein affinity prediction.

Various computational methods for binding affinity prediction have been proposed based on free energy perturbation, empirical scoring, and force-field potentials(Audie and Scarlata, 2007; Bogan and Thorn, 1998; Chothia and Janin, 1975; Horton and Lewis, 1992; Ma et al., 2002; Qin et al., 2011; Su et al., 2009). These scoring function based methods are typically trained and evaluated on limited datasets. These methods fail to accurately predict binding affinities for diverse datasets(Kastritis and Bonvin, 2010). Among computational binding affinity prediction methods, machine learning is preferred because of its implicit treatment of all factors involved in protein-protein interactions (PPIs) and the flexibility of using empirical data instead of a fixed or predetermined function form(Ain et al., 2015). A representation of the design and use of machine learning models for binding affinity prediction is given in Fig. 1. Machine learning based affinity prediction models require a dataset of diverse protein complexes with experimentally determined affinity values for training. By extracting the feature representation of protein complexes, a regression model is trained which can be used for affinity prediction of a novel complex (Fig. 1). A number of machine learning based studies for protein binding affinity prediction have been proposed in the literature(Moal et al., 2011; Tian et al., 2012; Vangone and Bonvin, 2015; Yugandhar and Gromiha, 2014). All of these studies are based on protein binding affinity benchmark dataset with 3-D structures of 144 protein complexes(Kastritis et al., 2011). The affinity prediction models proposed by Moal *et al.,* Tian *et al.,* and Vangone and Bonvin in their studies are based on 3-D protein structures(Moal et al., 2011; Tian et al., 2012; Vangone and Bonvin, 2015). However, protein structures are not available for most protein complexes. Consequently, sequence-based prediction of binding affinity is an important research problem. Sequence-based binding affinity prediction is challenging because proteins interaction and binding affinity are dependent upon protein structures and functions.

Among sequence-based protein binding affinity prediction models, the model proposed by Yugandhar and Gromiha (PPA-Pred2) is the state of the art predictor (Yugandhar and Gromiha, 2014). PPA-Pred2 claims high accuracy with a high correlation score between true and predicted binding affinity values. However, their proposed model performed poorly on an external validation dataset (Moal and Fernández-Recio, 2014). Furthermore, their prediction errors are, surprisingly, lower than the reported deviation in experimental measurements of binding affinity values and the error rates of structure-based prediction techniques (Kastritis et al., 2011; Moal and Fernández-Recio, 2014). Yugandhar and Gromiha have attributed this issue to the difference in experimental conditions and computational platforms (Yugandhar and Gromiha, 2015). In this work, we have replicated the validation of PPA-Prep2 on external independent test dataset as performed by Moal et. al (Moal and Fernández-Recio, 2014). These simple experiments have highlighted the need to revisit sequence-based binding affinity prediction and develop novel predictors that can be used in a practical setting. To address this, we have proposed a new binding affinity prediction model called ISLAND (In SiLico protein AffiNity preDictor). Our proposed model uses sequence features alone.

## 2. Methods

In this section, we discuss the details of our experimental design.

### 2.1. Datasets and preprocessing

We have used protein binding affinity benchmark dataset 2.0 for evaluation of PPA-Pred2 webserver and development of the proposed method ISLAND (Kastritis et al., 2011). This dataset contains 144 non-redundant complexes of proteins for which both bound and unbound structures of the ligand and receptor proteins are available. Protein binding affinities are given in terms of binding free energy ($\Delta G$) and disassociation constant ($K_d$). The binding free energy ($\Delta G$) ranges from -18.58 to -4.29. Following the same data curation and preprocessing technique used by Yugandhar and Gromiha, we have selected 135 complexes from this dataset (Yugandhar and Gromiha, 2014). This allows us to have a direct comparison of our method with the one proposed by Yugandhar and Gromiha (Yugandhar and Gromiha, 2014).

We have also used an external independent test dataset of 39 protein-protein complexes with known binding free energy (ΔG to perform a stringent test of performance comparison between PPA-Pred2 and ISLAND. This dataset is derived from Chen *et al.* by removing complexes having more than two chains, involving chains of size less than 50 residues, and having overlap with training data (Chen et al., 2013). This dataset has also been used by Moal *et. al.* in their evaluation of binding affinity prediction techniques (Moal and Fernández-Recio, 2014).

### 2.2. Evaluation of the PPA-Pred2 webserver

In order to investigate the accuracy of PPA-Pred2, we evaluated its performance on the selected dataset. For this purpose, we accessed PPA-Pred2 through its webserver (URL: http://www.iitm.ac.in/bioinfo/PPA_Pred/) on 03-02-2017 (Yugandhar and Gromiha, 2014). This webserver takes amino acid sequences of ligand and receptor of a protein complex and returns predicted values of change in binding free energy ($\Delta G$) and disassociation constant ($K_d$) (Yugandhar and Gromiha, 2014). The results obtained through this evaluation will also serve as baseline in this study. The predicted values obtained from the server are available as supplementary file (see "Supplementary Information").

### 2.3. Sequence homology as affinity predictor

In order to confirm whether simple homology is enough to predict protein binding affinity accurately or not, we have developed a sequence homology based protein binding affinity predictor as a baseline. For this purpose, we predicted the affinity value of a query protein complex based on the affinity value of its closest homolog in our dataset of protein complexes with known binding affinity values. We performed the Smith-Waterman alignment to determine the degree of homology between two protein complexes using BLOSUM-62 substitution matrix with gap opening and extension penalties of -11 and -1, respectively (Eddy, 2004; Smith and Waterman, 1981).

### 2.4. Proposed methodology

We have developed a sequence-only regression model called ISLAND (In SiLico protein AffiNity preDictor), for protein binding affinity prediction. To develop ISLAND, we have used different regression methods, evaluation protocols, and sequence-based feature extraction techniques. The methodology adopted for the development of ISLAND is detailed below.

### 2.5. Sequence-based features

In machine learning based prediction models, we require a feature representation of each example for training and testing (Fig. 1). Therefore, we have represented each complex in our dataset through a feature representation obtained from individual chains in the ligand ($l$) and receptor ($r$) of each complex. We used a number of explicit features and various kernel representations to model sequence based attributes of protein complexes. We discuss the sequence based features used in this study below.

#### 2.5.1. Explicit features

- *Amino Acid Composition features (AAC)*

These features capture the occurrences of different amino acids in a protein sequence. It gives a 20-dimensional feature vector $\boldsymbol{\phi}_{AAC}(\boldsymbol{s})$ of a given sequence $\boldsymbol{s}$ such that the $\boldsymbol{\phi}_{AAC}(\boldsymbol{s})_k$ contains the number of times amino acid $\boldsymbol{k}$ occurs in $\boldsymbol{s}$.(Leslie et al., 2002) This feature representation has successfully been used to predict protein interactions, binding sites and prion activity (Leslie et al., 2002; Minhas et al., 2017; Minhas and Ben-Hur, 2012).

- *Average BLOSUM-62 features (Blosum)*

In contrast to AAC, this feature representation models the substitutions of physicochemically similar amino acids in a protein. In this feature representation, protein sequence $\boldsymbol{s}$ is converted into a 20-dimensional feature vector by simply averaging the columns from the BLOSUM-62 substitution matrix corresponding to the amino acids in the given sequence. Mathematically, $\phi_{Blosum}(s) = \frac{1}{|s|}\sum_{i=1}^{|s|} B_i$, where $B_i$ is the column of the BLOSUM-62 substitution matrix (Eddy, 2004) corresponding to the i<sup>th</sup> residue in $s$.

- *Propy features (propy)*

In order to capture the biophysical properties of amino acids and sequence-derived structural features of a given protein sequence, we used a feature extraction package called propy (Cao et al., 2013). It gives a 1,537-dimensional feature representation $\phi_{propy}(s)$ of a given sequence $s$. This representation includes pseudo-amino acid compositions (PseAAC), autocorrelation descriptors, sequence-order-coupling number, quasi-sequence-order descriptors, amino acid composition, transition and the distribution of various structural and physicochemical properties (Limongelli et al., 2015; Li et al., 2006).

- *Position Specific Scoring Matrix features (PSSM)*

This feature representation models the evolutionary relationships between proteins. To get this representation, we used Position Specific Scoring Matrix (PSSM) of a given protein sequence (Altschul et al., 1997). We obtained the PSSM for each protein chain in a complex by using PSI-BLAST for three iterations against the non-redundant (nr) protein database with an e-value threshold of $10^{-3}$ (Altschul et al., 1997; Pruitt et al., 2005). In this feature representation, we represent the protein sequence $s$ by the average of columns in its PSSM. This results in a 20-dimensional feature vector $\phi_{PSSM}(s) = \frac{1}{|s|}\sum_{i=1}^{|s|} F_i^s$, where $F_i^s$ is the column in the PSSM corresponding to the i<sup>th</sup> residue in $s$.

- *ProtParam features (ProtParam)*

In order to capture different physiochemical properties of a protein such as molecular weight of the protein, aromaticity, instability index, isoelectric point, and secondary structure fractions, we have used ProParam ExPASy tools to get ProtParam representation (Gasteiger et al., 2005; Guruprasad et al., 1990; Lobry and Gautier, 1994). This leads to a 7-dimensional feature representation $\phi_{ProtParam}(s)$ of a given sequence $s$.

*2.5.2. Kernel representations*

In addition to using explicit protein sequence features in our machine learning models for binding affinity prediction, we have also experimented with different sequence based kernel (Ben-Hur and Noble, 2005; Cortes et al., 2008). Kernel methods present an alternate way of sequence representation by modeling the degree of similarity between protein sequences instead of an explicit feature representation (Ben-Hur and Noble, 2005). Kernel based methods such as support vector machines and support vector regression can make use of these kernel function scores in their training and testing (Ben-Hur et al., 2008). Different sequence kernels used in this work are described below. Each of these kernels $k(a, b)$ can be interpreted as a function that measures the degree of similarity between sequences $a$ and $b$.

- *Smith-Waterman alignment kernel (SW kernel)*

We have used the Smith-Waterman alignment algorithm for determining the degree of similarity between two protein sequences (Smith and Waterman, 1981). The Smith-Waterman kernel $k_{SW}(a,b)$ is simply the alignment score obtained from the Smith-Waterman local alignment algorithm using BLOSUM-62 substitution matrix with gap opening and extension penalties of -11 and -1, respectively. It is important to note that this kernel may not satisfy the Mercer's conditions as the eigen values of the kernel matrix may be negative.(Mercer, 1909) We addressed this issue by subtracting the most negative eigen value of the original kernel matrix from its diagonal elements (Saigo et al., 2004). From a theoretical point of view, this kernel can be interpreted as the optimal local alignment score of the two sequences (Saigo et al., 2004). Mathematically, the Smith-Waterman alignment score $k_{SW}(a,b)$ between sequences $a$ and $b$ can be written as follows (Saigo et al., 2004).

$$k_{SW}(a,b) = max_{\pi \in \Pi(l,r)} p(a,b,\pi) \qquad (1)$$

Here, $\Pi(a,b)$ denote the set of all possible local alignments between $a$ and $b$, and $p(a,b,\pi)$ represents the score of the local alignment $\pi \epsilon \Pi(a,b)$ between $a$ and $b$.

- *Local alignment kernel (LA kernel)*

Local alignment kernel is useful for comparing sequences of different lengths that share common parts (Ben-Hur et al., 2008; Saigo et al., 2004). In contrast to the Smith-Waterman alignment kernel which considers only the optimal alignment, this kernel sums up contributions of all the possible local alignments of input sequences. Mathematically, the local alignment score $k_{LA}(a,b)$ between sequences $a$ and $b$ can be written as follows (Saigo et al., 2004).

$$k_{LA}^{\beta}(a,b) = \sum_{\pi \in \Pi(a,b)} \exp(\beta p(a,b,\pi)) \qquad (2)$$

Here in Eq. (2), $\beta \geq 0$ is a parameter that controls the sensitivity of LA kernel. For larger values of $\beta$ score of LA kernel approaches SW kernel score (Saigo et al., 2004). We have used $\beta = 0.1$.

- *Mismatch kernel (MM kernel)*

The mismatch kernel captures the degree of overlap between subsequences of the two sequences while allowing mismatches (Leslie et al., 2004). MM kernel $k_{MM}^{k,m}(a,b)$ gives the number of subsequences of length $k$ that are present in both the input sequences $a$ and $b$ with a maximum of $m$ mismatches. Ranges for the values of $k$ and $m$ are $3-9$ and $0-5$, respectively. We have used $k=5$ and $m=3$.

### 2.6. Complex level features representation

We need to predict protein binding affinity at the complex level. Since we have extracted features at the chain level, therefore, we require a mechanism to obtain a complex level feature representation from individual chains. The basic mechanism of combining individual chain level feature representation from each ligand and receptor to form a complex level representation is shown in Fig. 2. Complex level representation is obtained for explicit features by concatenation of chain level features and for kernels by adding kernels over the constituent chains of a complex.

### 2.6.1. Feature concatenation

In our machine learning model, a complex $c$ is represented by the tuple $c \equiv ((l,r), y)$, where $(l, r)$ is the pair of ligand and receptor proteins in the complex and $y$ is the corresponding affinity value. To generate the complex level feature representation $\boldsymbol{\psi}(c)$, we simple concatenate the feature representations of respective ligand and receptor as $\boldsymbol{\psi}(c) = \begin{bmatrix} \boldsymbol{\psi}_{Avg}(l) \\ \boldsymbol{\psi}_{Avg}(r) \end{bmatrix}$. Here, $\boldsymbol{\psi}_{Avg}(l) = \frac{1}{|l|}\sum_{q \in l}\boldsymbol{\phi}(q)$ and $\boldsymbol{\psi}_{Avg}(r) = \frac{1}{|r|}\sum_{q \in r}\boldsymbol{\phi}(q)$ are the explicit feature representations averaged across all the chains present in the ligand and receptor proteins, respectively. This method of feature representation generation has already been used for protein interacting residues predictor (Ahmad and Mizuguchi, 2011).

### 2.6.2. Combining kernels

To make predictions at the complex level from sequence based kernels, we have developed a complex-level kernel by simply averaging the kernel function values of individual chains from the two complexes (Ben-Hur and Noble, 2005). Mathematically, the kernel over complexes $c$ and $c'$ is given by $K(c, c') = \frac{1}{|c| \times |c'|}\sum_{q \in c, q' \in c'} k(q, q')$, where $k(q, q')$ is the chain level kernel over two chains from the two complexes.

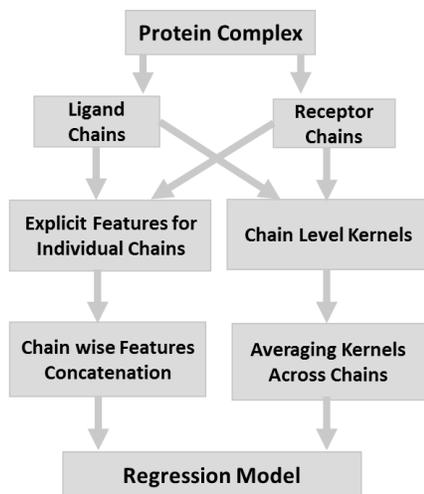

**Fig. 2.** The techniques adopted for generating feature representation of a protein complex for machine learning.

## 2.7. Regression models

Here, we begin by presenting the binding affinity prediction problem as a regression problem. In machine learning based affinity prediction, a dataset consisting of $N$ examples $(c_i, y_i)$, where $i = 1 \dots N$. In this representation, $c_i$ is a complex with known binding affinity $y_i$.

The feature representation of $c_i$ is $\boldsymbol{\psi}(c_i)$. Our objective in machine learning based regression is to train a model $f(c)$ that can predict the binding affinity of the complex $c$. The learned regression function $f(\cdot)$ should generalize well over previously unseen complexes. We used the following regression techniques through Scikit-learn to get different regression models (Pedregosa et al., 2011). It is also important to note that the feature representations are normalized to have unit norm and standardized to zero mean and unit standard deviation before using them in the regression model.

### 2.7.1. Ordinary Least-Squares Regression (OLSR)

Ordinary least squares (OLS) estimates the regression function $f(c) = \boldsymbol{w}^T \boldsymbol{\psi}(c) + b$ by minimizing the sum of squared error between the actual and predicted affinity values $\min_{w,b} \sum_i^N (y_i - f(\boldsymbol{c}_i))^2$ (Watson, 1967). Here, $\boldsymbol{w}$ and $b$ are parameters to be learned. This technique has been used previously for protein binding affinity prediction.(Yugandhar and Gromiha, 2014) We have used this technique as a baseline in our study.

### 2.7.2. Support Vector Regression (SVR)

Support Vector Machines have been effectively used to solve different computational problems in bioinformatics (Cortes and Vapnik, 1995). Support Vector regression (SVR) performs regression using $\varepsilon$-insensitive loss and, by controlling model complexity (Smola and Schölkopf, 2004). Training a SVR for protein binding affinity prediction involves optimizing the objective function given in Eq. (3) to learn a regression function $f(c) = \boldsymbol{w}^T \boldsymbol{\psi}(c) + b$.

$$min_{w,b} \frac{1}{2} \|\boldsymbol{w}\|^2 + C \sum_{i=1}^{N} (\xi_i^+ + \xi_i^-)$$
$$Such\ that\ for\ all\ i: \begin{cases} y_i - f(\boldsymbol{c}_i) \leq \varepsilon + \xi_i^+ \\ f(\boldsymbol{c}_i) - y_i \leq \varepsilon + \xi_i^- \\ \xi_i^+, \xi_i^- \geq 0 \end{cases} \quad (3)$$

Here, $\frac{1}{2}\|\boldsymbol{w}\|^2$ controls the margin, $\xi_i^+$ and $\xi_i^-$ capture the extent of margin violation for a given training example and $C$ is the penalty of such violations (Cortes and Vapnik, 1995). We used both linear and radial basis function (rbf) SVR in this study. The values of C, gamma, and epsilon were optimized during model selection. SVR has already been used for the same purpose in previous studies .(Yugandhar and Gromiha, 2014)

### 2.7.3. Random Forest Regression (RFR)

Random Forest regression (RFR) is an ensemble of regression trees used for nonlinear regression (Breiman, 2001). Each regression tree in the RF is based on randomly sampled subsets of input features. We optimized RF with respect to the number of decision trees and a minimum number of samples required to split in this study. This regression technique has been used in many related studies (Ballester and Mitchell, 2010; Li et al., 2014; Moal et al., 2011).

**Table 1.** Evaluation of PPA-Pred2 through its webserver on affinity benchmark dataset 2.0.

| Group | Total Complexes | Through PPA-Pred2 webserver | | | Reported in (Yugandhar and Gromiha, 2014) |
|---|---|---|---|---|---|
| | | $P_r$ | P-value | RMSE | $P_r$ |
| Antigen-Antibody | 15 | -0.36 | 0.205 | 2.55 | 0.854 |
| Enzyme-Inhibitor | 31 | 0.52 | 0.006 | 2.47 | 0.739 |
| Other enzymes | 20 | 0.06 | 0.860 | 4.93 | 0.765 |
| G-protein containing | 16 | 0.60 | 0.114 | 5.26 | 0.953 |
| Receptor containing | 12 | -0.40 | 0.257 | 5.79 | 0.931 |
| Non-cognate | 09 | Option not available in web interface | | | 0.986 |
| Miscellaneous1 | 11 | 0.64 | 0.086 | 2.13 | 0.992 |
| Miscellaneous2 | 10 | -0.77 | 0.074 | 2.08 | 0.983 |
| Miscellaneous3 | 11 | 0.44 | 0.274 | 2.42 | 0.980 |
| **Combined** | **126** | **0.43** | **1.8e-05** | **3.61** | |

## 2.8. Model validation and performance assessment

To evaluate the performance of all the trained regression models, we have used Leave One Complex Out (LOCO) cross-validation (CV) (Abbasi and Minhas, 2016). In LOCO, a regression model is developed with $(N-1)$ complexes and tested on the left out complex. This process is repeated for all the $N$ complexes present in the dataset. We used Root Mean Squared Error RMSE $= \sqrt{\frac{1}{n}\sum_{i=1}^{N}(y_i - f(c_i))^2}$ and Pearson correlation coefficient $(P_r)$ between the predicted $f(c_i)$ and actual $y_i$, as performance measures for model evaluation and performance assessment. To check the statistical significance of the results, we have also estimated P-value of the correlation coefficient scores. We used grid search over training data to find the optimal values of hyper-parameters of different regression models.

## 3. Results and Discussion

In this section, we present and discuss the results and major outcomes of our study.

### 3.1. Binding affinity prediction through sequence homology

As a baseline, we have obtained the predicted affinity values of all 135 complexes in our dataset using sequence homology based affinity prediction method. The Pearson correlation coefficient $(P_r)$ between predicted and experimental values of $\Delta G$ is 0.29 with a Root Mean Squared Error ($RMSE$) of 3.20. These results with poor correlation and high RMSE value show that the sequence homology only cannot be effectively used to predict binding affinity of the protein complexes. As discussed in the next section, our machine learning based method performs significantly better than homology based predictions.

### 3.2. Binding affinity prediction through ISLAND

We have evaluated the performance of three different regression models (OLSR, RFR, and SVR) along eight different types of sequence descriptors with LOCO cross-validation over the docking benchmark dataset. The results of this analysis are shown in Table 2 in the form of Root Mean Squared Error ($RMSE$) and Pearson correlation coefficient ($P_r$) along with statistical

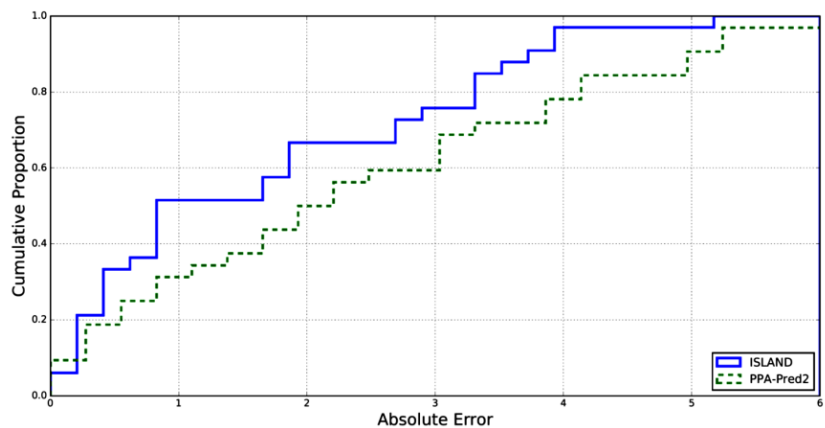

**Fig. 3.** Cumulative histogram of absolute error between actual and predicted binding affinity values through ISLAND and PPA-Pred2 on validation dataset

significance (P-value). With explicit features, we obtained a maximum correlation of 0.41 with $RMSE = 2.60$ between predicted and experimental values of $\Delta G$ using propy through SVR (Table 2). While using kernel descriptors, we obtained a maximum correlation of 0.44 with a $RMSE = 2.56$ between predicted and experimental $\Delta G$ values using the local alignment kernel (Table 2). We have achieved the best performance through local kernel across all sequence descriptors used in this study (Table 2). Moreover, LA kernel performs better than SW kernel because of considering the effect of all the local alignments rather taking best alignment as in case of SW kernel. The $RMSE$ value of ISLAND predictions is quite close to the range of experimental uncertainties (1-2 kcal/mol) as reported by Kastritis et al(Kastritis et al., 2011).

The performance of ISLAND is also comparable with the methods based on 3-D protein structures such as DFIRE ($P_r = 0.35$), PMF ($P_r = 0.37$), RBF ($P_r = 0.44$), M5' ($P_r = 0.45$) and RF ($P_r = 0.48$) as reported by Moal et al. (Moal et al., 2011). In spite of getting the comparable performance of ISLAND with structure-based methods, there is still a lot of room for the improvement in affinity prediction from sequence information alone.

### 3.3. Comparison on external independent test dataset

We obtained the predicted binding affinity values for all the complexes in our external validation dataset using both PPA-Pred2 and ISLAND. We have seen a significant performance improvement of ISLAND in terms of RMSE between predicted and experimental $\Delta G$ values. We obtained an RMSE of 2.20 with ISLAND whereas PPA_Pred2 gives us an RMSE of 3.62. We have find a comparable performance of both the methods in terms of Pearson correlation coefficient. We have also shown a comparison between ISLAND and PPA-Pred2 in terms of absolute error between predicted and actual binding affinity values of all the complexes in our validation set in Fig. 3. The binding affinity of >50% complexes were predicted within an absolute error of 1.5 kcal/mol using ISLAND, whereas, through PPA-Pred2 absolute error for these complexes is above 2.5kcal/mol (see Fig. 3). These results show better performance of our proposed method for binding affinity prediction of proteins in a complex in comparison to PPA-Pred2. Moreover, these results also support the criticism of Moal *et. at*., on PPA-Pred2 and suggests a need for further work on methods of protein binding affinity prediction using sequence information (Moal and Fernández-Recio, 2014).

## 4. Conclusions and future work

This paper highlights the fact that the true generalization performance of even the state of the art sequence-only predictor of binding affinity is far from satisfactory and that the development of effective and practical methods in this domain is still an open problem. We also propose a novel sequence-only predictor of binding affinity called ISLAND which gives better accuracy than PPA-Pred2 webserver and other existing methods over the same validation set.

**Conflict of interest**

We have no conflict of interest to declare.

**Acknowledgments**

The authors are thankful to K. Yugandhar and M. Michael Gromiha, Indian Institute of Technology Madras, India for providing relevant data for this study. We would also like to thank Dr. Hanif Durad and Dr. Javaid Khurshid, DCIS, PIEAS, Pakistan for helping us meet the computational requirements of the project. We also acknowledge the very fruitful discussions with Dr. Asa Ben-Hur, Colorado State University, Fort Collins, USA over the course of this project.

**Funding**

Wajid A. Abbasi is supported by a grant (PIN: 213-58990-2PS2-046) under indigenous 5000 Ph.D. fellowship scheme from the Higher Education Commission (HEC) of Pakistan.

**References**

Abbasi, W.A., Minhas, F.U.A.A., 2016. Issues in performance evaluation for host–pathogen protein interaction prediction. J. Bioinform. Comput. Biol. 14, 1650011. doi:10.1142/S0219720016500116

Ahmad, S., Mizuguchi, K., 2011. Partner-Aware Prediction of Interacting Residues in Protein-Protein Complexes from Sequence Data. PLOS ONE 6, e29104. doi:10.1371/journal.pone.0029104

Ain, Q.U., Aleksandrova, A., Roessler, F.D., Ballester, P.J., 2015. Machine-learning scoring functions to improve structure-based binding affinity prediction and virtual screening. WIREs Comput Mol Sci 5, 405–424. doi:10.1002/wcms.1225

Alberts, B., Johnson, A., Lewis, J., Raff, M., Roberts, K., Walter, P., 2002. Protein Function.

Altschul, S.F., Madden, T.L., Schäffer, A.A., Zhang, J., Zhang, Z., Miller, W., Lipman, D.J., 1997. Gapped BLAST and PSI-BLAST: a new generation of protein database search programs. Nucleic Acids Res 25, 3389–3402.

Audie, J., Scarlata, S., 2007. A novel empirical free energy function that explains and predicts protein–protein binding affinities. Biophysical Chemistry 129, 198–211. doi:10.1016/j.bpc.2007.05.021

Ballester, P.J., Mitchell, J.B.O., 2010. A machine learning approach to predicting protein-ligand binding affinity with applications to molecular docking. Bioinformatics 26, 1169–1175. doi:10.1093/bioinformatics/btq112

Ben-Hur, A., Noble, W.S., 2005. Kernel methods for predicting protein–protein interactions. Bioinformatics 21, i38–i46. doi:10.1093/bioinformatics/bti1016

Ben-Hur, A., Ong, C.S., Sonnenburg, S., Schölkopf, B., Rätsch, G., 2008. Support Vector Machines and Kernels for Computational Biology. PLoS Comput. Biol. 4, e1000173. doi:10.1371/journal.pcbi.1000173

Bogan, A.A., Thorn, K.S., 1998. Anatomy of hot spots in protein interfaces. J. Mol. Biol. 280, 1–9. doi:10.1006/jmbi.1998.1843

Breiman, L., 2001. Random Forests. Mach Learn 45, 5–32. doi:10.1023/A:1010933404324

Cao, D.-S., Xu, Q.-S., Liang, Y.-Z., 2013. propy: a tool to generate various modes of Chou's PseAAC. Bioinformatics 29, 960–962. doi:10.1093/bioinformatics/btt072

Chen, J., Sawyer, N., Regan, L., 2013. Protein–protein interactions: General trends in the relationship between binding affinity and interfacial buried surface area. Protein Sci 22, 510–515. doi:10.1002/pro.2230

Chothia, C., Janin, J., 1975. Principles of protein–protein recognition. Nature 256, 705–708. doi:10.1038/256705a0

Cortes, C., Mohri, M., Rostamizadeh, A., 2008. Learning sequence kernels, in: 2008 IEEE Workshop on Machine Learning for Signal Processing. Presented at the 2008 IEEE Workshop on Machine Learning for Signal Processing, pp. 2–8. doi:10.1109/MLSP.2008.4685446

Cortes, C., Vapnik, V., 1995. Support-Vector Networks. Machine Learning 20, 273–297. doi:10.1023/A:1022627411411

Eddy, S.R., 2004. Where did the BLOSUM62 alignment score matrix come from? Nat Biotech 22, 1035–1036. doi:10.1038/nbt0804-1035

Gasteiger, E., Hoogland, C., Gattiker, A., Duvaud, S. 'everine, Wilkins, M., Appel, R., Bairoch, A., 2005. Protein Identification and Analysis Tools on the ExPASy Server, in: Walker, J. (Ed.), The Proteomics Protocols Handbook. Humana Press, pp. 571–607.

Guruprasad, K., Reddy, B.V., Pandit, M.W., 1990. Correlation between stability of a protein and its dipeptide composition: a novel approach for predicting in vivo stability of a protein from its primary sequence. Protein Eng. 4, 155–161.

Horton, N., Lewis, M., 1992. Calculation of the free energy of association for protein complexes. Protein Sci 1, 169–181.

Kastritis, P.L., Bonvin, A.M.J.J., 2013. On the binding affinity of macromolecular interactions: daring to ask why proteins interact. J R Soc Interface 10, 20120835. doi:10.1098/rsif.2012.0835

Kastritis, P.L., Bonvin, A.M.J.J., 2010. Are scoring functions in protein-protein docking ready to predict interactomes? Clues from a novel binding affinity benchmark. J. Proteome Res. 9, 2216–2225. doi:10.1021/pr9009854

Kastritis, P.L., Moal, I.H., Hwang, H., Weng, Z., Bates, P.A., Bonvin, A.M.J.J., Janin, J., 2011. A structure-based benchmark for protein-protein binding affinity. Protein Sci. 20, 482–491. doi:10.1002/pro.580

Kessel, A., Ben-Tal, N., 2010. Introduction to Proteins: Structure, Function, and Motion. CRC Press, Boca Raton, FL, USA.

Leslie, C., Eskin, E., Noble, W.S., 2002. The spectrum kernel: a string kernel for SVM protein classification. Pac Symp Biocomput 564–575.

Leslie, C.S., Eskin, E., Cohen, A., Weston, J., Noble, W.S., 2004. Mismatch string kernels for discriminative protein classification. Bioinformatics 20, 467–476. doi:10.1093/bioinformatics/btg431

Li, H., Leung, K.-S., Wong, M.-H., Ballester, P.J., 2014. Substituting random forest for multiple linear regression improves binding affinity prediction of scoring functions: Cyscore as a case study. BMC Bioinformatics 15, 291. doi:10.1186/1471-2105-15-291

Limongelli, I., Marini, S., Bellazzi, R., 2015. PaPI: pseudo amino acid composition to score human protein-coding variants. BMC Bioinform. 16, 123. doi:10.1186/s12859-015-0554-8

Li, Z.R., Lin, H.H., Han, L.Y., Jiang, L., Chen, X., Chen, Y.Z., 2006. PROFEAT: a web server for computing structural and physicochemical features of proteins and peptides from amino acid sequence. Nucl. Acids Res. 34, W32–W37. doi:10.1093/nar/gkl305

Lobry, J.R., Gautier, C., 1994. Hydrophobicity, expressivity and aromaticity are the major trends of amino-acid usage in 999 Escherichia coli chromosome-encoded genes. Nucleic Acids Res 22, 3174–3180. doi:10.1093/nar/22.15.3174

Ma, X.H., Wang, C.X., Li, C.H., Chen, W.Z., 2002. A fast empirical approach to binding free energy calculations based on protein interface information. Protein Eng. 15, 677–681.

Mercer, J., 1909. Functions of Positive and Negative Type, and their Connection with the Theory of Integral Equations. Philosophical Transactions of the Royal Society of London A: Mathematical, Physical and Engineering Sciences 209, 415–446. doi:10.1098/rsta.1909.0016

Minhas, F. ul A.A., Ben-Hur, A., 2012. Multiple instance learning of Calmodulin binding sites. Bioinformatics 28, i416–i422. doi:10.1093/bioinformatics/bts416

Minhas, F. ul A.A., Ross, E.D., Ben-Hur, A., 2017. Amino acid composition predicts prion activity. PLOS Computational Biology 13, e1005465. doi:10.1371/journal.pcbi.1005465

Moal, I.H., Agius, R., Bates, P.A., 2011. Protein-protein binding affinity prediction on a diverse set of structures. Bioinformatics btr513. doi:10.1093/bioinformatics/btr513

Moal, I.H., Fernández-Recio, J., 2014. Comment on "protein–protein binding affinity prediction from amino acid sequence." Bioinformatics btu682. doi:10.1093/bioinformatics/btu682

Pedregosa, F., Varoquaux, G., Gramfort, A., Michel, V., Thirion, B., Grisel, O., Blondel, M., Prettenhofer, P., Weiss, R., Dubourg, V., Vanderplas, J., Passos, A., Cournapeau, D., Brucher, M., Perrot, M., Duchesnay, É., 2011. Scikit-learn: Machine Learning in Python. J. Mach. Learn. Res. 12, 2825−2830.

Pruitt, K.D., Tatusova, T., Maglott, D.R., 2005. NCBI Reference Sequence (RefSeq): a curated non-redundant sequence database of genomes, transcripts and proteins. Nucleic Acids Res 33, D501–D504. doi:10.1093/nar/gki025

Qin, S., Pang, X., Zhou, H.-X., 2011. Automated prediction of protein association rate constants. Structure 19, 1744–1751. doi:10.1016/j.str.2011.10.015

Saigo, H., Vert, J.-P., Ueda, N., Akutsu, T., 2004. Protein homology detection using string alignment kernels. Bioinformatics 20, 1682–1689. doi:10.1093/bioinformatics/bth141

Smith, T.F., Waterman, M.S., 1981. Identification of common molecular subsequences. J. Mol. Biol. 147, 195–197. doi:10.1016/0022-2836(81)90087-5

Smola, A.J., Schölkopf, B., 2004. A tutorial on support vector regression. Statistics and Computing 14, 199–222. doi:10.1023/B:STCO.0000035301.49549.88

Su, Y., Zhou, A., Xia, X., Li, W., Sun, Z., 2009. Quantitative prediction of protein-protein binding affinity with a potential of mean force considering volume correction. Protein Sci. 18, 2550–2558. doi:10.1002/pro.257

Tian, F., Lv, Y., Yang, L., 2012. Structure-based prediction of protein-protein binding affinity with consideration of allosteric effect. Amino Acids 43, 531–543. doi:10.1007/s00726-011-1101-1

Tomlinson, I.M., 2004. Next-generation protein drugs. Nat Biotech 22, 521–522. doi:10.1038/nbt0504-521

Vangone, A., Bonvin, A.M., 2015. Contacts-based prediction of binding affinity in protein–protein complexes. eLife 4, e07454. doi:10.7554/eLife.07454

Watson, G.S., 1967. Linear Least Squares Regression. Ann. Math. Statist. 38, 1679–1699. doi:10.1214/aoms/1177698603

Wilkinson, K.D., 2004. Quantitative analysis of protein-protein interactions. Methods Mol. Biol. 261, 15–32. doi:10.1385/1-59259-762-9:015

Yugandhar, K., Gromiha, M.M., 2015. Response to the comment on "protein- protein binding affinity prediction from amino acid sequence." Bioinformatics 31, 978–978. doi:10.1093/bioinformatics/btu821

Yugandhar, K., Gromiha, M.M., 2014. Protein-protein binding affinity prediction from amino acid sequence. Bioinformatics 30, 3583–3589. doi:10.1093/bioinformatics/btu580